\documentclass[twocolumn,aps,prl,showpacs,superscriptaddress,floatfix]{revtex4}
\usepackage{dcolumn}
\usepackage{amsmath}
\usepackage{bm}

\begin{document}

\title{Two--Loop Bethe Logarithms}

\author{Krzysztof Pachucki}
\affiliation{Institute of Theoretical Physics, Warsaw University,
ul.~Ho\.{z}a 69, 00--681 Warsaw, Poland}

\author{Ulrich D.~Jentschura}
\affiliation{National Institute of Standards and Technology,
Gaithersburg, MD20899-8401}

\begin{abstract}
We calculate the two-loop Bethe logarithm correction to atomic
energy levels in hydrogen-like systems. The two-loop Bethe
logarithm is a low-energy quantum electrodynamic (QED) effect
involving multiple summations over virtual excited atomic states.
Although much smaller in absolute magnitude than the well-known
one-loop Bethe logarithm, the two-loop analog is quite significant
when compared to the current experimental accuracy of the
$1S$--$2S$ transition: it contributes -8.19 and -0.84 kHz for the
$1S$ and the $2S$ state, respectively. The two-loop Bethe
logarithm has been the largest unknown correction to the hydrogen
Lamb shift to date. Together with the ongoing measurement of the
proton charge radius at the Paul Scherrer Institute its
calculation will bring theoretical and experimental accuracy for
the Lamb shift in atomic hydrogen to the level of $10^{-7}$.
\end{abstract}


\maketitle

\noindent In 1947 Hans Bethe explained the splitting of $2S_{1/2}$
and $2P_{1/2}$ levels in hydrogen by the presence of the electron
self-interaction \cite{bethe}, and expressed it in terms of the
``Bethe'' logarithm. For S states this quantity may be represented
as a matrix element involving the logarithm of the nonrelativistic
Hamiltonian of the hydrogen atom. In natural units with $\hbar = c
= \epsilon_0 = 1$ and $m$ denoting the electron mass, it reads
\begin{equation}
\label{bl1l} \ln k_0(nS) = \frac{\left\langle \vec p \, (H-E) \,
\ln\left[ \frac{\displaystyle 2(H-E)} {\displaystyle (Z\alpha)^2
\,m} \right]\vec p\,\right\rangle}{ \left\langle \vec p \,
(H-E)\vec p\,\right\rangle}\,.
\end{equation}
This Bethe logarithm is due to the emission and subsequent
absorption of a single soft virtual photon (it is independent of
the nuclear charge number $Z$ and depends only on the principal
quantum number $n$ and the orbital angular momentum which is zero
for $S$ states). Over the years, QED theory has been developed and
refined \cite{eides}, and various additional radiative,
relativistic, and combined corrections have been obtained to face
the increasing precision of the measurements of the hydrogen
spectrum \cite{gar, paris}. These include higher-order
relativistic one--, two--, and three--loop corrections, nuclear
recoil, finite-size corrections, and even the nuclear polarizability. The
modern all-order calculation of the leading one-loop self-energy
was developed by Mohr in \cite{mohr} and significantly improved
recently using convergence acceleration techniques which led to a
highly accurate evaluation of the fully relativistic Green
function \cite{ulj3}. One of the conceptually most difficult and
as well as interesting corrections involve 
nuclear recoil effects. The finite
nuclear mass, although large as compared to the electron mass,
prohibits the use of the one-body Dirac equation, and alternative
approaches such as the Bethe-Salpeter equation or nonrelativistic
QED \cite{lepage} have been introduced. Although these methods are quite
general, no compact formulas have been derived for relativistic
recoil effects. A few years ago, Shabaev tackled the problem of
recoil corrections to hydrogenic energy levels of first order in
the mass ratio, deriving expressions which are nonperturbative in the
nuclear charge (see a recent review in~\cite{shab}), and this has
led to the current highly accurate calculations of relativistic recoil
corrections.

Another class of effects, namely binding two--loop corrections,
are quite difficult from a numerical point of view. A detailed
investigation of these effects has been performed only in the 
last years. The nonperturbative treatment (no $Z\alpha$-expansion) of
the two-loop bound-state corrections has been pursued by various
groups in Refs.~\cite{person,sap,lab,yer}. However, these
calculations were mostly performed for high-$Z$ hydrogen-like
atoms. As yet, complete results have not been obtained for $Z=1$
(see for example the most recent work in \cite{yer}). In the
perturbative treatment of the bound-state two-loop self-energy
correction, one calculates terms in a semi-analytic expansion in
$Z \alpha$ and $\ln[(Z \alpha)^{-2}]$. For $S$ states, the first
nonvanishing terms read
\begin{eqnarray}
\Delta E &=& \left(\frac{\alpha}{\pi}\right)^2 \,
\frac{(Z\alpha)^4}{n^3} \, H(Z\alpha)\, m\,, \label{02} \\[2ex]
H(Z\alpha) &=& B_{40} + (Z\alpha)\, B_{50}
+ (Z\alpha)^2 \, \bigl\{
B_{63} \, \ln^3(Z\alpha)^{-2} + \nonumber \\ & &
B_{62} \, \ln^2(Z\alpha)^{-2} + B_{61} \, \ln(Z\alpha)^{-2} + B_{60} \bigr\}
\nonumber \\ &&+\ldots \,. \label{03}
\end{eqnarray}
It was a perhaps surprising result, found only
a few years ago~\cite{krp3}, that this expansion has a very slow convergence:
because of the large absolute magnitude of higher-order coefficients,
many terms have to be included for reliable theoretical
predictions which in order to match the current 
experimental precision. One of the
remaining unknown but relevant contributions is the two-loop Bethe
logarithm, which forms the dominant part of the problematic
nonlogarithmic coefficient $B_{60}$ (note that theoretical effort
in evaluating the one-loop analog of this coefficient, $A_{60}$,
has extended over more than three
decades~\cite{eri1,eri2,a60,krp2}). The two-loop Bethe logarithm
originates from the emission and absorption of two virtual soft
photons. Using nonrelativistic QED one derives the following
expression for this two-loop correction \cite{krp}.
For convenience we set $Z=1$, pull out a common prefactor
and express remaining in terms of dimensionless quantities:
\begin{equation}
\label{2}
\Delta E = \left( \frac{2 \alpha}{3 \pi} \right)^2 \,\alpha^6\,m
\int^{\epsilon_1}_0 {\rm d}\omega_1
\int^{\epsilon_2}_0 {\rm d}\omega_2 \,
f(\omega_1, \omega_2) \,,
\end{equation}
where 
\begin{eqnarray}
f(\omega_1, \omega_2) &=&\omega_1\omega_2\bigl\{
\left< p^i \, G(\omega_1) \, p^j \,
G(\omega_1 + \omega_2) \, p^i \,
G(\omega_2) \, p^j \right>  \nonumber\\[1ex]
& & + \left< p^i \, G(\omega_1) \, p^j \,
G(\omega_1 + \omega_2) \, p^j \,
G(\omega_1) \, p^i \right>/2 \nonumber\\[1ex]
& & + \left< p^i \, G(\omega_2) \, p^j \,
G(\omega_1 + \omega_2) \, p^j \,
G(\omega_2) \, p^i \right>/2
\nonumber\\[1ex]
& & +
\left< p^i \, G(\omega_1) \, p^i \,
G'(0) \, p^j \,
G(\omega_2) \, p^i \right>
\nonumber\\[1ex]
& & - \left< p^i \, G(\omega_1) \, p^i \right> \,
\left< p^j \, G^2(\omega_2) \, p^i \right>/2
\nonumber\\[1ex]
& & - \left< p^i \, G(\omega_2) \, p^i \right> \,
\left< p^j \, G^2(\omega_1) \, p^i \right>/2
\nonumber\\[1ex]
& & - \left< p^i \, G(\omega_1) \,
G(\omega_2) \, p^i \right>
\nonumber\\[1ex]
& &
- \left< p^i \, [G(\omega_1)+G(\omega_2)] \,
p^i \right>/(\omega_1 + \omega_2)\bigr\} \,.\label{05}
\end{eqnarray}
Here, $p^k = -i\partial^k$,
$G(\omega) =  1/[E - (H + \omega)]$ is the nonrelativistic
Green function, $E=-1/(2 n^2)$ is the Schr\"{o}dinger energy of the reference
state with $H=\vec p^2/2-1/r$, and $G'(0) = 1/(E - H)'$ is the reduced
Green function with the reference state excluded. The
$\omega$-integrals in Eq. (\ref{2}) depend on $\epsilon_1$ and
$\epsilon_2$. We are free to choose the relation between
the parameters $\epsilon_1$ and $\epsilon_2$. Following~\cite{krp},
we perform the expansion in large $\epsilon_2$ first, and next in
large $\epsilon_1$. All the terms involving $1/\epsilon_1$ or
$1/\epsilon_2$ are neglected. The linear and logarithmic terms,
using the same relations, have already been considered in
\cite{krp}. The dependence on the $\epsilon$-parameters cancels 
when the contributions from the high- and the low-energy photons are added. 
The constant term, which by definition we call the 
two-loop Bethe logarithm, is calculated numerically here.

There are two ways to calculate the integrand in Eq.~(\ref{05}).
The first one relies on the use of known analytic expressions for the
Schr\"odinger-Coulomb Green function, which involve the product of 
Whittaker functions. The precise calculation of the $\omega$-integrals
requires the use of large $\omega_1$ and $\omega_2$. This leads to
a number of problems, including a numerical overflow in the calculation
of these functions which persists even in quadruple precision,
and this approach has therefore not been pursued here.
In the second way, which is chosen here, the Schr\"odinger
Hamiltonian is represented on a numerical grid, as a large
symmetric band matrix \cite{sal}. Each inversion in Eq. (\ref{05})
corresponds to a solution of a linear equation with a known
right-hand side. This process is quite fast since it scales
linearly with the number of grid points and is numerically stable. For the
final evaluation we used $10^5$ grid points, and we have checked
the numerical accuracy of the results against those obtained with
$2\cdot10^5$ grid points.

Having calculated the matrix elements, we proceed to the evaluation of
the $\omega_1$- and $\omega_2$-integrals.
In accordance with the definition of two-loop Bethe logarithm,
we first fix $\omega_1$, integrate over $\omega_2$ and
expand in large $\epsilon_2$:
\begin{eqnarray}
f(\omega_1) &=& \int_0^{\epsilon_2}
d\omega_2\,f(\omega_1,\omega_2) \nonumber \\ &=&
\epsilon_2\,a(\omega_1) +\ln(\epsilon_2)\,b(\omega_1) +
g(\omega_1) \,. \label{06}
\end{eqnarray}
As in the case of the one-loop Bethe logarithm, the $g$ function finds a
representation that is suited for a numerical computation,
\begin{equation}
\label{evalg}
g(\omega_1) = {\cal I}_1 + {\cal I}_2 + {\cal I}_3\,,
\end{equation}
where
\begin{subequations}
\begin{eqnarray}
\label{I1}
{\cal I}_1 &=& \int_0^M {\rm d}\omega_2 \, f(\omega_1, \omega_2) \,,
\\[2ex]
\label{I2}
{\cal I}_2 &=& \int_M^\infty {\rm d}\omega_2 \,
\left[ f(\omega_1, \omega_2) - a(\omega_1) -
\frac{b(\omega_1)}{\omega_2} \right]\,,
\\[2ex]
\label{I3}
{\cal I}_3 &=& a(\omega_1) \, M + b(\omega_1) \, \ln M\,,
\end{eqnarray}
\end{subequations}
with arbitrary $M$. The $a$ and $b$ coefficients are the first terms
of the expansion of $f(\omega_1, \omega_2)$ for large
$\omega_2$ at fixed $\omega_1$,
\begin{eqnarray}
f(\omega_1, \omega_2) &=& a(\omega_1) +
\frac{b(\omega_1)}{\omega_2} +
\frac{2\,\sqrt{2}\,b(\omega_1)}{\omega_2^{3/2}} +
\nonumber \\[2ex] &&
c(\omega_1)\,\frac{\ln(\omega_2)}{\omega_2^2} +
\frac{d(\omega_1)}{\omega_2^2} + \ldots
\label{09}
\end{eqnarray}
The first coefficient is
\begin{equation}
a(\omega_1) = \omega_1\,\left< p^i \, \frac{H-E}{(H-E+\omega_1)^2}
\, p^i \right> \,,
\end{equation}
and the second reads
\begin{equation}
b(\omega_1) = \omega_1 \,\, \delta_{\pi\delta^3(r)} \left\{ \left<
p^i \, \frac{1}{E-(H+\omega_1)}\, p^i \right>  \right\} \,,
\end{equation}
where by $\delta_V$ we denote the first-order correction to the
specified matrix elements. Namely, $\phi$, $E$, and $H$ receive
corrections according to
\begin{subequations}
\begin{eqnarray}
H &\to& H + V\,, \\[2ex]
|\phi\rangle &\to& |\phi\rangle +
\frac{1}{(E-H)'}\,V|\phi\rangle\,, \\[2ex]
E &\to& E + \langle V \rangle \,.
\end{eqnarray}
\end{subequations}
The higher-order coefficients $c,d,\ldots$ in Eq.~(\ref{09}) are
obtained from the fit to the numerical data and are subsequently
used for the analytic integration at large $\omega_2$. The
numerical integration over $\omega_2$ is performed with a well
adapted set of 400 grid points, and the accuracy is checked by
comparison with a calculation involving 200 grid points. Results
of this integration for few chosen values of $\omega_1$ is shown
in Table \ref{table1}.
\begin{table}
\begin{minipage}{8.6cm}
\caption{Sample values of the $g$ function, defined in Eq.~(\ref{06}),
 for the $1S$ and $2S$ states.} 
\label{table1}
\begin{tabular}{rrr}
\hline
\hline
$\omega$ & $g_{1S}$ & $8\,g_{2S}$ \\
\hline
   0  &   0.000\,00	&     0.000\,00	\\
   5  & -10.281\,60	&   -10.367\,94	\\
  20  & -16.560\,34	&   -16.415\,97	\\
  80  & -22.714\,02	&   -22.439\,66	\\
 180  & -26.232\,35	&   -25.923\,09	\\
 320  & -28.699\,64	&   -28.376\,26	\\
 500  & -30.599\,22	&   -30.268\,75	\\
 720  & -32.142\,95	&   -31.808\,43	\\
\hline
\hline
\end{tabular}
\end{minipage}
\end{table}
The next step is the numerical integration over $\omega_1$,
\begin{equation}
\Delta E = \left( \frac{2 \alpha}{3 \pi} \right)^2 \,\alpha^6\,m\,
\int^{\epsilon_1}_0 {\rm d}\omega_1 \,g(\omega_1) \,. \label{13}
\end{equation}
In order to perform this integration, one needs to know the
large-$\omega_1$ asymptotics of $g$, which for an $nS$ state
reads,
\begin{eqnarray}
\label{14}
g(\omega_1) &=&\biggl\{
- 4 \ln\omega_1 + 2\,\left[\ln 2 - 1 - \ln k_0(nS)\right]
\nonumber\\[2ex]
& & + \frac{4 \, \sqrt{2}}{\sqrt{\omega_1}} \,
\left[\ln(\omega_1) + 2 \, (\ln 2 - 1) - \pi \right] +
\nonumber\\[2ex]
& & + \frac{1}{\omega_1} \,
\left[
\ln^2(\omega_1) + 8 + \frac{3}{2} \, N(nS) + 5 \pi^2
\right]
\nonumber\\[2ex]
& & +{\mathcal A}\,\frac{\ln^2(\omega_1)}{\omega^{3/2}_1}
    +{\mathcal B}\,\frac{\ln(\omega_1)}{\omega^{3/2}_1}
    +{\mathcal C}\,\frac{1}{\omega^{3/2}_1} \nonumber \\ &&
+ {\mathcal O}\left(\frac{1}{\omega^2_1}\right)\biggr\}\,
\frac{1}{n^3}\,.
\end{eqnarray}
Here, $N$ denotes a nonlogarithmic in $\alpha$ correction to the
Bethe logarithm induced by a Dirac $\delta$. It has been calculated
in~\cite[Eq.~(12)]{ulj1}, and the results for the $1S$ and $2S$
states read
\begin{subequations}
\begin{eqnarray}
N(1S) &=& 17.855\,672(1)\,,\\
N(2S) &=& 12.032\,209(1)\,.
\end{eqnarray}
\end{subequations}
The terms proportional to the ${\mathcal A},{\mathcal B},{\mathcal
C}$-coefficients in Eq.~(\ref{14}), and the omitted higher-order
terms are obtained from the fit to the calculated data. The
numerical stability of the parameters obtained from the fit, in
various ranges of $\omega_1$, indicates consistency of the
numerically determined values for $g(\omega_1)$ with the
analytically derived logarithmic terms in Eq.~(\ref{14}). This
constitutes a check for the large value for the coefficient
$B_{61}$ derived in \cite{krp} on the basis of the logarithmic
asymptotics.

The integral in Eq. (\ref{13}) is performed in analogy to the
algorithm presented in Eq. (\ref{06}), by choosing an arbitrary
value for the parameter $M$ (e.g., $M=1$), and dropping all linear and
logarithmic terms in $\epsilon_1$. For $\omega_1$ larger than 720,
the extrapolated values from the fit are employed. However, this
part of the integral depends significantly on the unknown analytic 
behavior of $g(\omega)$ at large $\omega$, and this is
the main source of the integration uncertainty. The
overall result of the numerical integration leads to the following
nonlogarithmic terms of the order of $(\alpha/\pi)^2 \,
(Z\alpha)^6/n^3$ relative to the electron rest mass, whose
numerical coefficients we choose to denote by $b_L$,
\begin{subequations}
\begin{eqnarray}
b_L(1S) &=& -81.4(3)\,, \\
b_L(2S) &=& -66.6(3)\,.
\end{eqnarray}
\end{subequations}
These terms are much larger than the corresponding one-loop Bethe
logarithms for typical hydrogenic states, but suppressed in
absolute magnitude by an additional factor
$\alpha\,(Z\alpha)^2/\pi$. For hydrogen ($Z=1$), the above results
contribute -8.19 and -0.84 kHz to the $1S$ and $2S$ states,
respectively. The other contributions to $B_{60}$ are considered
below; the notation is consistent with that of Ref. \cite{krp}.
The coefficient $B_{60}$ can be represented as the sum
\begin{equation}
B_{60} = b_L+b_{M}+b_{F}+b_{H}+b_{\rm VP}\,.
\end{equation}
The two-loop Bethe logarithm $b_L$
comes from the region where both photon momenta are small
and has been the subject of this work.
$b_{M}$ stems from an integration region
where one momentum is large $\sim m$, and the second
momentum is small.
This contribution is given by a Dirac $\delta$ correction to
the Bethe logarithm. It
has already been derived in \cite{krp} but not included
in the theoretical predictions for the Lamb shift:
\begin{equation}
b_{M} = \frac{10}{9}\,N(nS)\;
\end{equation}
$b_{F}$ and $b_{H}$ originate from a region where both photon
momenta are large $\sim m$, and the electron momentum is small and
large respectively. Finally, $b_{\rm VP}$ is a contribution from
diagrams that involve a closed fermion loop. None of these effects
have been calculated as yet. On the basis of our experience with
the one- and two-loop calculations \cite{remark}
we estimate the magnitude of
these uncalculated terms to be of the order of 15\%. This leads to
the following overall result for the $B_{60}$ coefficients:
\begin{subequations}
\begin{eqnarray}
B_{60}(1S) &=& -61.6(3)\pm 15\% \label{21}\,, \\
B_{60}(2S) &=& -53.2(3)\pm 15\% \label{22}\,,
\end{eqnarray}
\end{subequations}
and to the following corrections to transition frequencies
\begin{subequations}
\begin{eqnarray}
\delta \nu(1S) &=& -6.20(93)\,{\rm kHz}\,, \\
\delta \nu(2S) &=& -0.67(10)\,{\rm kHz}\,.
\end{eqnarray}
\end{subequations}
In the foreseeable future, we may expect to have
results from a direct numerical calculation of the two-loop
self-energy at low $Z$.
In addition to improving our knowledge of the Lamb shift at low $Z$,
the obtained result will then serve as a
consistency check between two different approaches to bound-state
QED. 

With the results obtained in Eqs. (\ref{21},\ref{22}) we are
in a position to present theoretical predictions for the Lamb
shifts of $1S$ and $2S$ states. Based on the former result
obtained in \cite{krp} and the corrections calculated in this
work, we obtain:
\begin{subequations}
\begin{eqnarray}
\nu_L(1S) &=& 8\,172\,811(32)(2)\,{\rm kHz}\,, \\
\nu_L(2S) &=& 1\,045\,005(4)\,{\rm kHz}\,,
\end{eqnarray}
\end{subequations}
where the first error comes from the
current uncertainty in the proton charge radius $r_p$,
and the second one is a rough estimate of uncalculated
terms: $b_{H}, b_{\rm VP}$, as well as
higher-order two-loop corrections denoted
by dots in Eq. (\ref{03}), and the three-loop
binding correction $C_{50}$~(for a recent evaluation of
$C_{40}$ see Ref.~\cite{mr}). Theoretical predictions for the $1S$ state
agree well with the experimental value of the combined result from
the Garching and the Paris groups~\cite{gar,paris,val}:
\begin{equation}
\nu_L(1S)_{\rm exp} = 8\,172\,840(22)\,{\rm kHz}\,.
\end{equation}
Since the uncertainty coming from the proton structure dominates
theoretical uncertainties for the Lamb shift in hydrogen, an
experiment at the Paul Scherrer Institute (PSI) is currently being
pursued to make a precise measurement of $r_p$ from
the Lamb shift in muonic hydrogen \cite{muonic}. This system is
much more sensitive to the proton charge radius. Once it is
measured, the combined hydrogen and muonic Lamb shift will test
QED at a precision level of $10^{-7}$.

\acknowledgments
This work was supported in part by the research grant from
European Commission under contract HPRI-CT-2001-50034.
U. D. J. wishes to acknowledge helpful
conversations with P. J. Mohr and J. Sims.


\begin{thebibliography}{00}
\bibitem{remark}
Concerning logarithmic two-loop vacuum-polarization effects
\cite{krp}: The contribution of the two-loop self-energy diagrams
to $B_{61}$ for the $1S$ state reads $49.7$, and the diagrams
that involve a closed fermion loop amount to $0.6$. Concerning
the one-loop higher-order binding correction $A_{60}(1S)$ (analog
of $B_{60}$) as an example: The result for $1S$ is $-30.92415(1)$
(see Refs.~\cite{krp2,ulj3}). This is the sum of a contribution
due to low-energy virtual photons of
$-27.3$~\cite[Eq.~(5.116)]{krp2}, and a high-energy term of about
$-3.7$~\cite[Eq.~(6.102)]{krp2}.
\end{thebibliography}
\end{document}